\documentclass[prc,twocolumn,superscriptaddress,superscriptfootnote]{revtex4-1}
\usepackage{amssymb}
\usepackage[toc,page]{appendix}
\usepackage{subcaption}
\usepackage{graphicx}
\usepackage{caption}
\usepackage{amsmath,amsfonts,amssymb,epsfig,color}
\usepackage{multirow}
\newcommand{\ket}[1]{\ensuremath{\left| #1 \right\rangle}}

\begin{document}

\title{Exact isovector pairing in a shell-model framework:\\ Role of proton-neutron correlations in isobaric analog states}
\author{M. E. Miora}
\affiliation{Department of Physics,
Rollins College, Winter Park, FL 32789, USA}
\affiliation{Department of Physics and Astronomy, Louisiana State University,
Baton Rouge, LA 70803, USA}
\author{K. D. Launey}
\affiliation{Department of Physics and Astronomy, Louisiana State University,
Baton Rouge, LA 70803, USA}
\author{D. Kekejian}
\affiliation{Department of Physics and Astronomy, Louisiana State University,
Baton Rouge, LA 70803, USA}
\author{F. Pan}
\affiliation{Department of Physics and Astronomy, Louisiana State University,
Baton Rouge, LA 70803, USA}
\affiliation{Department of Physics,
Liaoning Normal University, Dalian 116029, People's Republic of China}
\author{J. P. Draayer}
\affiliation{Department of Physics and Astronomy, Louisiana State University,
Baton Rouge, LA 70803, USA}

\date{\today}

\begin{abstract}

\noindent We utilize a nuclear shell model Hamiltonian with only two adjustable parameters to generate, for the first time, exact solutions for pairing correlations for light to medium-mass nuclei, including the challenging proton-neutron pairs, while also identifying the primary physics involved. In addition to single-particle energy and Coulomb potential terms, the shell model Hamiltonian consists of an isovector $T=1$ pairing interaction and an average proton-neutron isoscalar $T=0$ interaction, where the $T=0$ term describes the average interaction between non-paired protons and neutrons. This Hamiltonian is exactly solvable, where, utilizing 3 to 7 single-particle energy levels, we reproduce experimental data for 0$^+$ state energies for isotopes with mass $A=10$ through $A=62$ exceptionally well including isotopes from He to Ge. Additionally, we isolate effects due to like-particle and proton-neutron pairing, provide estimates for the total and proton-neutron pairing gaps, and reproduce $N$ (neutron) = $Z$ (proton) irregularity. These results provide a further understanding for the key role of proton-neutron pairing correlations in nuclei, which is especially important for waiting-point nuclei on the rp-path of nucleosynthesis.

\end{abstract}

\maketitle

\section{Introduction}
Since a pairing model was first applied to nuclei by Bohr, Mottelson, and Pines \cite{BohrMottelsonPines}, studies have repeatedly found pairing correlations to have a profound influence on nuclear structure \cite{Belyaev}. A better understanding of pairing features in nuclei could greatly benefit other areas of research, such as superfluidity in neutron stars \cite{BennemannHeinz,GezerlisPethick}, pairing correlations in nuclear matter \cite{DrischlerKruger,DingRios,BurrelloColonna} and nuclei around closed shells \cite{KDL85}. While pairing correlations among like-particles, e.g., proton-proton ($pp$) and neutron-neutron ($nn$) pairing, have been described through numerous methods \cite{Helmers,Flowers,Lane,Hecht1,HendersonDukelsky} and are well understood, proton-neutron ($pn$) pairing has been less studied due to its complexity \cite{Hecht2,VanIsacker,PanDraayer,SviratchevaGD_PRC04,Georgiev,PAN201886,PAN20181}. For example, current approaches for pairing in the continuum have been addressed \cite{MercenneDukelsky,Rodolfo1,Rodolfo2} but solely for like-particle pairing. An accurate treatment of the challenging $pn$ pairing interaction has been suggested to be important for understanding waiting-point nuclei in rapid-proton capture nucleosynthesis \cite{Sharma,Sharma2,PruetFuller} and may play a role in neutrinoless double-beta decay ($0\nu \beta \beta$) \cite{HinoharaEngel,JiaoEngel}. 
Therefore, exact analytic solutions for both like-particle and $pn$ pairing are of great interest. 

\

Albeit restricted to degenerate single-particle energies, exact solutions to like-particle and $pn$ pairing interactions can be achieved through the $T=1$ charge-independent pairing Hamiltonian constructed using generators of the quasispin group Sp$_j$(4), where $j$ indicates the orbits utilized in the model space and $T$ corresponds to the isospin \cite{Hecht2, LauneyThesis}. For non-degenerate single-particle energies, approximate numerical solutions can be attained through the BCS formalism \cite{Amand,ErreaDukelsky,Baranger,Bredmond,FlowersVujicic}. 
Some studies utilize the algebraic Bethe ansatz method with an infinite-dimensional Lie algebra \cite{DukelskyEsebbag,PanDraayerOrmand,PanDraayer2,PanDraayer3,PanDraayerGuo,Richardson5,ChenRichardson,HTChenRichardson} and other methods \cite{Richardson1,Richardson2,Richardson3,Richardson4,DukelskyLerma,ChongChen,Dukelsky} provide exact solutions for systems with like-particle pairing or for systems with two or fewer pairs. 

\

In this paper, we present a new shell model Hamiltonian that yields exact analytic solutions for the lowest \textit{isovector-paired} $0^+$ states for up to six nucleons (three pairs). The Hamiltonian, adapted from Ref. \cite{SviratchevaGD_PRC04} where degenerate energies have been considered, consists of a single-particle energy term, Coulomb potential term, and includes an isovector $T=1$ pairing interaction and an isoscalar $T=0$ proton-neutron interaction that accounts for the average interaction between non-paired nucleons. The model utilizes the analytic solutions to isovector pairs in non-degenerate single-particle levels that are derived in Ref. \cite{PanDraayer} for up to three pairs. However, when considering three or more pairs highly nonlinear equations appear and require sophisticated solution mechanisms \cite{GuanLauney}. Here, we report applications of such solutions to light through medium-mass nuclei including the challenging $pn$ pairs. We also identify the primary physics involved through an analysis of the staggering behavior of our results and pairing gap estimates. 

\section{Theoretical Formalism}
Algebraic solutions to a $T=1$ charge-independent pairing Hamiltonian that utilizes single-particle energies of the $j$th orbit, $\epsilon_{j}$, which can be derived from the spherical shell model, are introduced in Ref. \cite{PanDraayer}. These solutions are for $J^{\pi}=0^{+}$ states of $2k$ nucleons and include both like-particle and $pn$ pairs, where $k$ is the total number of pairs. 
To describe ground states and $0^{+}$ isobaric analog states in nuclei it is important to consider the Coulomb potential and an isoscalar $T=0$ $pn$ interaction \cite{SviratchevaGD_PRC04} in addition to the isovector pairing. In particular, our model Hamiltonian is expressed as

\begin{eqnarray}
\label{model}
\hat{H}=\sum_{j} \varepsilon_{j}N_{j}&-&G\sum_{jj^{'}\mu} A^{\dagger}_{j,\mu}A_{j^{'},\mu}\\ \nonumber
&+&\alpha \Bigg(\hat{T}^2-\frac{N}{2}\bigg(\frac{N}{2}+1 \bigg) \Bigg)
+V_{\text{Coul}},
\end{eqnarray}

where $G>0$ is the pairing strength, and $\alpha$ is the strength of the additional isoscalar $T=0$ interaction, 
which can be understood as the average interaction between nonpaired protons and neutrons in a $T=1$ pair as shown in Ref. \cite{LauneyThesis} (also related to the symmetry term).
The nucleon number operator $N_{j}$, pair creation $A^{\dagger}_{j,\mu}$ and annihilation $A_{j,\mu}=(A^{\dagger}_{j,\mu})^\dagger$ operators, where $\mu=+,-,0$ indicates $pp$, $nn$, and $pn$ pairs, respectively, together with the isospin operators $\hat T_{j,\pm 1}$ and $\hat T_{j,0}$ are generators of the Sp$_{j}$(4) group. The total number operator is given by $N=\sum_j N_j$, which is also $N=2k$, and $V_{\text{Coul}}$ denotes the Coulomb interaction. The Hamiltonian is initially solved for the first two terms in Eq. (\ref{model}),  as described in the next section, resulting in eigenstates that have $k$, $T$, and $T_z$  as good quantum numbers (or, equivalently, proton and neutron numbers along with $T$); in this basis the $\alpha$ term is diagonal, resulting in additional energy given by  $\alpha (T(T+1)-k(k+1))$. The Coulomb term is also diagonal and its contribution is accounted by an estimate given in Ref. \cite{Retamosa}, as described in Sec. \ref{Vc}.

\subsection{Exact isovector pairing solutions for up to six nucleons \label{PanDT1}}
Exact solutions for non-degenerate single-particle energies and isovector $T=1$ pairing interaction [first two terms of Eq. (\ref{model})] for $k\leq 3$ are derived for selected 
permutations of the permutation group $S_k$ in Ref. \cite{PanDraayer}. As described in Ref. \cite{PanDraayer}, the method uses elements of the Gaudin algebra $\mathcal{G}$(Sp(4)), $A^{\dagger}_{\mu}(x)=\sum_j \frac{A^{\dagger}_{j,\mu}}{1+\varepsilon_{j}x}$, $A_{\mu}(x)=\sum_j \frac{A_{j,\mu}}{1+\varepsilon_{j}x}$, $T_{\mu}(x)=\sum_j \frac{T_{j,\mu}}{1+\varepsilon_{j}x}$, and $N(x)=\sum_j \frac{N_{j,\mu}}{1+\varepsilon_{j}x}$, where $x=\{x_1,x_2,\dots,x_k\}$ are spectral parameters for $k$ pairs. Hence, one can solve the Hamiltonian, $\hat{H}=\left. \partial N/ \partial  x \right |_{x=0} +G A^\dagger(0) \cdot A(0)$ using the Bethe ansatz wave function $\ket{k;\zeta; [\lambda]_k,TT_z}=\sum_{P \in S_k} Q^{[\lambda]}(x_{P_1},x_{P_2},\dots, x_{P_k})$ $\left\{  A^\dagger(x_{P_1}) \times A^\dagger(x_{P_k}) \times \dots  \times A^\dagger(x_{P_k}) \right\}^{TT_z}\ket{0}$, that describes a $k$-paired state with $\ket{0}$ the seniority-zero state, where $[\lambda]_k$ is an irrep of the permutation group $S_k$ containing $k$ boxes in the corresponding Young diagram and $P$ labels all possible permutations. As a result, the expansion coefficients $Q^{[\lambda]}$ and the spectral parameters, $x_1,\dots,x_k$, are determined. In Ref. \cite{PanDraayer} solutions are derived for the cases $k=1$, $T=1$ along with $k=2$ and $3$ for $T=0, \dots, k$.

\

It is important to note that this method leads to highly nonlinear equations that become more challenging to solve as $k$ increases. Therefore, to find solutions and reduce the number of singularities we have modified the spectral parameters of Ref. \cite{PanDraayer}, such that in numerical calculations we use $y_i\equiv 2/x_i$ where $i=1,2,\dots, k$. Additionally, we utilize an average single-particle energy $\epsilon_{\text{avg}}$ defined as 

\begin{equation}
\label{eavg}
\epsilon_{avg}=\frac{\sum\limits_{j}(\Omega_{j} \epsilon_{j})}{\sum\limits_{j} \Omega_{j}},
\end{equation}
where $\Omega_{j}=j+\frac{1}{2}$ is the $j$-level degeneracy. Hence, the energies utilized are taken with respect to this average energy, $\varepsilon_{j}=\epsilon_{j}-\epsilon_{avg}$. In the appendix we briefly outline  the main equations, which have been derived in Ref. \cite{PanDraayer}, in terms of  the different variables used in the present numerical calculations.

\subsection{Coulomb potential \label{Vc}}
We include the Coulomb potential ($V_{\text{Coul}}$) by using estimates provided in Ref. \cite{Retamosa}.  Defining $N_{+}$, $N_{-}$, and $A$ as the valence proton, neutron, and the atomic numbers of nuclei, respectively, we first calculate $V_{\text{Coul}}$ of isotopes with $N_{+}=N_{-}$. These energies are then used to calculate $V_{\text{Coul}}$ when $Z>Z_{s}$ and $Z<Z_{s}$. If $N_{+}=N_{-}$ and $N_{+}\leq 20$ then $V_{\text{Coul}}$ is given as

\begin{equation}
\label{NZVcoul1}
V_{\text{Coul}}(A,N_{-})=0.162N_{-}^{2}+0.95N_{-}-18.25.
\end{equation}
However, if $N_{+}=N_{-}$ and $N_{+}> 20$ then $V_{\text{Coul}}$ is defined as

\begin{equation}
\label{NZVcoul2}
V_{\text{Coul}}(A,N_{-})=0.125N_{-}^{2}+2.35N_{-}-31.53.
\end{equation}

Next we calculate $V_{\text{Coul}}$ when $N_{+}\ne N_{-}$ and $N_{+}>N_{-}$, where the relation for this case is

\begin{eqnarray}
\label{Vcoul1}
V_{\text{Coul}}(A,N_{+})&=&  V_{\text{Coul}}(A,N_{+}-1) \\ \nonumber &+&1.44\frac{N_{+}-\frac{1}{2}}{A^{1/3}}-1.02.
\end{eqnarray}

However, if $N_{+}\ne N_{-}$ and $N_{+}<N_{-}$ then $V_{\text{Coul}}$ is

\begin{eqnarray}
\label{Vcoul2}
V_{\text{Coul}}(A,N_{+})&=& V_{\text{Coul}}(A,N_{+}+1) \\ \nonumber &-&1.44\frac{N_{+}+\frac{1}{2}}{A^{1/3}}+1.02.
\end{eqnarray}

The relations (\ref{NZVcoul1}-\ref{Vcoul2}) were used to calculate the Coulomb potentials for even-$A$ isotopes in the mass ranges $A=10-16$, $A=34-46$, and $A=50-56$. These energies were accounted for when reproducing the experimental energy spectra for these mass ranges.
\section{Results and Discussion}

The present model, which accounts for both $pn$ and like-particle pairing, has been successfully applied to even-$A$ nuclei for up to six particles above and below the $^{16}$O, $^{40}$Ca, and $^{56}$Ni cores. In particular, using only two adjustable parameters, $G$ and $\alpha$, and experimentally deduced single-particle energies we calculate exact solutions for the $J^{\pi}=0^{+}$ binding energies in even-even ($ee$) nuclei and the lowest isobaric analog $0^{+}$ excited states in odd-odd ($oo$) nuclei (which correspond to the ground state of the even-even neighbor), together with pair-excitation $0^+$ states. Using these solutions we are able to reproduce the experimental energy spectra as well as utilize discrete derivatives of the energy function to describe fine pairing features of these light to medium-mass nuclei off closed shells. 

\subsection{Energy spectra for isotopes around $^{\textbf{16}}$O, $^{\textbf{40}}$Ca, and $^{\textbf{56}}$Ni}

In our model we use single-particle energies deduced from the experimental energy spectrum of $A_{\text{core}} \pm 1$ nuclei for a core of mass $A_{\text{core}}$. These single-particle energies are non-degenerate, thus providing more accurate solutions as compared to an earlier algebraic model based on the Sp(4) group \cite{SviratchevaGD_PRC04}, or equivalently on the O(5) group, that utilizes the same Hamiltonian. Indeed, it is crucial for our model to consider non-degenerate energies due to the comparatively large energy gap between levels, which is on the order of approximately 1 MeV. The experimentally deduced single-particle energies and the model parameters utilized in the Hamiltonian are listed in Table \ref{spentable} for their respective cores. It should be noted that the $0d_{5/2}$ single-particle energy level in the $^{55}$Ni energy spectrum has yet to be experimentally determined, leaving us unable to utilize this energy in our Hamiltonian. This truncation in the model space may account for deviations from experiment for even-$A$ isotopes in the mass range $50 \le A \le 54$. 

\begin{figure*}[htbp]
\centering
\begin{subfigure}[t]{0.325\textwidth}
\includegraphics[width=\textwidth]{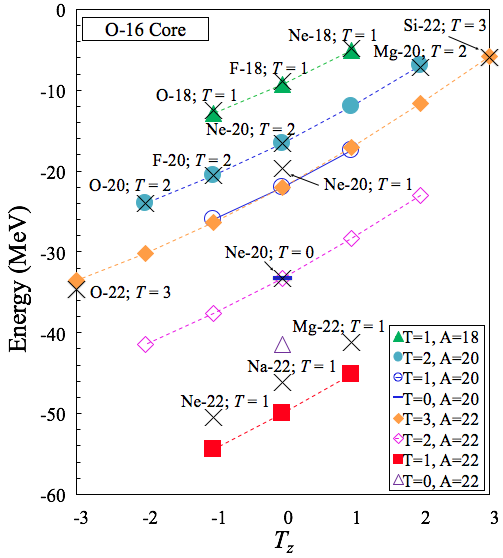}    
\caption{\centering}
\label{fig1}
\end{subfigure}
\begin{subfigure}[t]{0.325\textwidth}
\includegraphics[width=\textwidth]{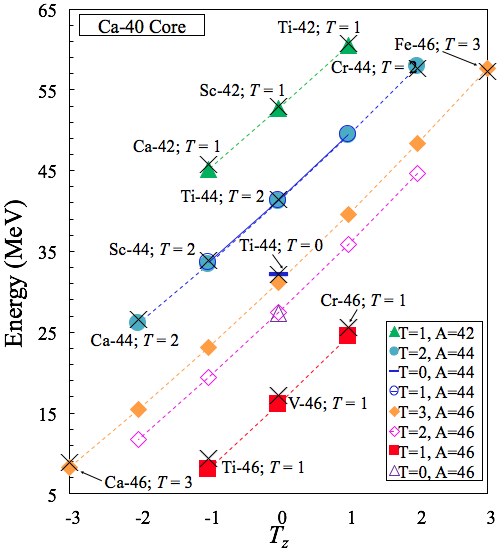}
\caption{\centering}
\label{fig2}
\end{subfigure}
\begin{subfigure}[t]{0.325\textwidth}
\includegraphics[width=\textwidth]{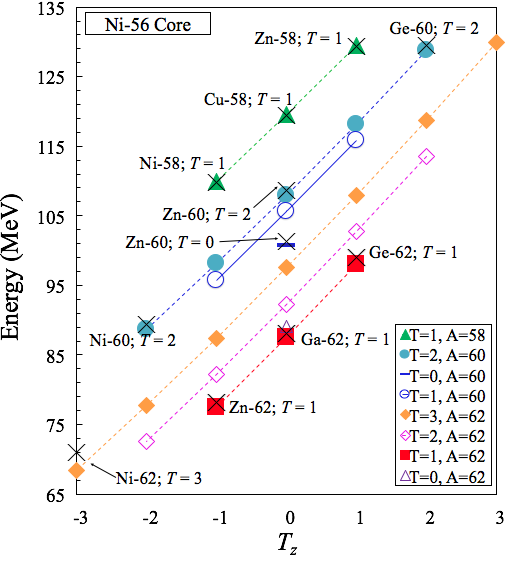}
\caption{\centering}
\label{fig3}
\end{subfigure}

\medskip
\begin{subfigure}[t]{0.325\textwidth}
\includegraphics[width=\textwidth]{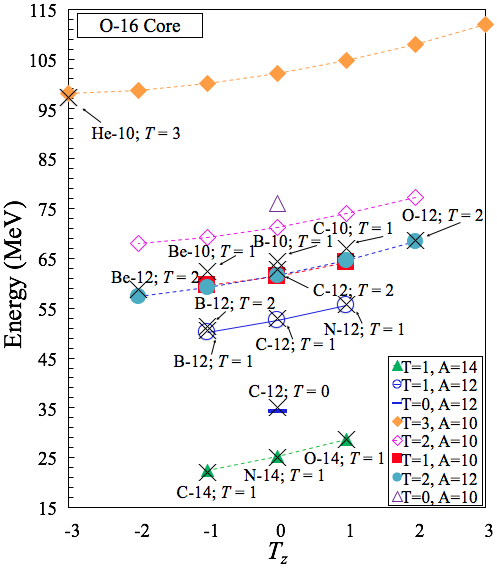}
\label{energyspectra}
\caption{\centering}
\end{subfigure}
\begin{subfigure}[t]{0.325\textwidth}
\includegraphics[width=\textwidth]{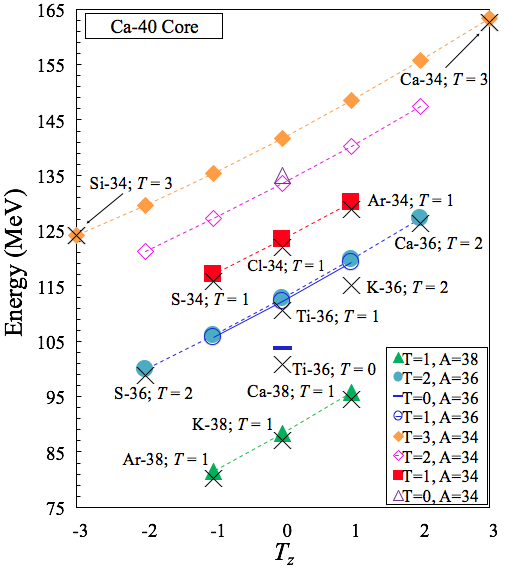}
\caption{\centering}
\label{fig2b}
\end{subfigure}
\begin{subfigure}[t]{0.33\textwidth}
\includegraphics[width=\textwidth]{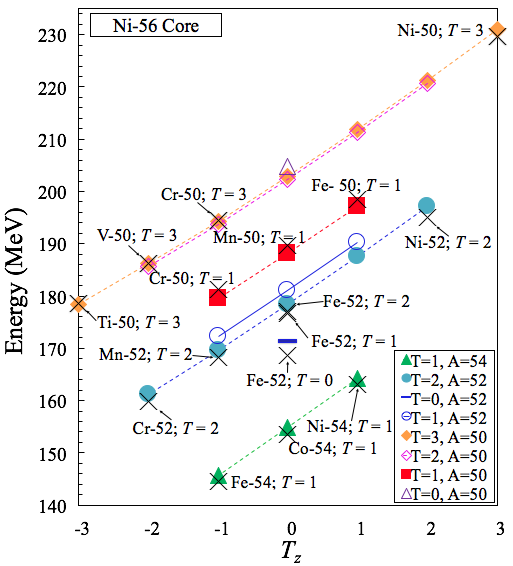}
\caption{\centering}
\label{fig3b}
\end{subfigure}
\caption{(Color online) Theoretical energy spectra (colored shapes) compared to experiment (black crosses) for $0^{+};0,..,3$ binding energies and lowest isobaric analog $0^{+};0,..,3$ excited states of isotopes above and below the (a) $^{16}$O core; (b) $^{40}$Ca core; (c) $^{56}$Ni core.}
\label{energyspectra}
\end{figure*}

\

The model parameters were determined by first adjusting $G$ for the $k=T$ cases where the $\alpha$-term of the Hamiltonian has zero contribution to the $0^+$ energy of the isobaric analog states. Next, $\alpha$ was adjusted for the $k\ne T$ cases where both the $pn$ isoscalar and isovector interactions contribute significantly. We find that, while typically particles and holes (above and below a core, respectively) can be described by the same $G$ and $\alpha$ values, a larger pairing strength, $G$, is required for the lightest nuclei below the $^{16}$O core in the mass range 10 $\leq$ $A$ $\leq$ 14. This, however, is in agreement with $G$ proportional to $(17\pm 1)/A $ and $\alpha$ proportional to $36\pm 3/A $, which is supporded by earlier estimates \cite{KDL85}. 

\

Our model very closely reproduces the energy of the lowest $0^{+},\ T=0, \dots,3$ states of $ee$ and $oo$ nuclei for up to six particles above and below the $^{16}$O, $^{40}$Ca, and $^{56}$Ni cores (Fig. \ref{energyspectra}). The theoretical and experimental energy spectrum of individual isotopes are listed for allowed isospin values. Though only like-particle pairing occurs when $k=|T_z|$,  our model accounts for $pn$ pairing as well, which is a significant feature, as it permits the calculation of the binding energies for isotopes when $k \neq |T_z|$ and the especially interesting $N=Z$ case. 
\
\begin{table}[]
\centering
\begin{tabular}{cc|c|}
\multicolumn{3}{l}{}                                                          \\ \hline
\multicolumn{1}{|c|}{Core}                       & \multicolumn{1}{c|}{Particles (MeV)} & \multicolumn{1}{c|}{Holes (MeV)} \\ \hline
\multicolumn{1}{|c|}{\multirow{9}{*}{$^{16}$O}}  & \multicolumn{1}{c|}{$G=0.55$}        & \multicolumn{1}{c|}{$G=1.65$}    \\
\multicolumn{1}{|c|}{}                           & \multicolumn{2}{c|}{$\alpha=2.445$}                                     \\ \cline{2-3} 
\multicolumn{1}{|c|}{}                           & $\varepsilon_{0d_{5/2}}=-4.14$           & $\varepsilon_{0p_{1/2}}=15.66$       \\
\multicolumn{1}{|c|}{}                           & $\varepsilon_{1s_{1/2}}=-3.27$           & $\varepsilon_{0p_{3/2}}=21.84$       \\
\multicolumn{1}{|c|}{}                           & $\varepsilon_{0p_{3/2}}=0.94$            & $\varepsilon_{0s_{1/2}}=23.22$       \\
\multicolumn{1}{|c|}{}                           & $\varepsilon_{0f_{7/2}}=1.55$            &                                  \\
\multicolumn{1}{|c|}{}                           & $\varepsilon_{1d_{3/2}}=0.41$            &                                  \\
\multicolumn{1}{|c|}{}                           & $\varepsilon_{0f_{5/2}}=-0.29$           &                                  \\
\multicolumn{1}{|c|}{}                           & $\varepsilon_{1p_{1/2}}=-1.09$           &                                  \\ \hline
\multicolumn{1}{|c|}{\multirow{7}{*}{$^{40}$Ca}} & \multicolumn{2}{c|}{$G=0.45$}                                           \\ 
\multicolumn{1}{|c|}{}                           & \multicolumn{2}{c|}{$\alpha = 1.229$}                                   \\ \cline{2-3} 
\multicolumn{1}{|c|}{}                           & $\varepsilon_{0f_{7/2}}=-8.36$           & $\varepsilon_{1s_{1/2}}=18.10$       \\
\multicolumn{1}{|c|}{}                           & $\varepsilon_{1p_{3/2}}=-6.42$           & $\varepsilon_{0d_{3/2}}=15.64$       \\
\multicolumn{1}{|c|}{}                           & $\varepsilon_{0f_{5/2}}=-5.79$           & $\varepsilon_{0d_{5/2}}=20.07$       \\
\multicolumn{1}{|c|}{}                           & $\varepsilon_{1p_{1/2}}=-4.75$           &                                  \\
\multicolumn{1}{|c|}{}                           & $\varepsilon_{0g_{9/2}}=-3.91$           &                                  \\ \hline
\multicolumn{1}{|c|}{\multirow{6}{*}{$^{56}$Ni}} & \multicolumn{2}{c|}{$G=0.33$}                                           \\ 
\multicolumn{1}{|c|}{}                           & \multicolumn{2}{c|}{$\alpha = 1.000$}                                   \\ \cline{2-3}
\multicolumn{1}{|c|}{}                           & $\varepsilon_{1p_{3/2}}=-10.25$          & $\varepsilon_{1s_{1/2}}=19.83$       \\
\multicolumn{1}{|c|}{}                           & $\varepsilon_{0f_{5/2}}=-9.48$           & $\varepsilon_{0d_{3/2}}=20.40$       \\
\multicolumn{1}{|c|}{}                           & $\varepsilon_{1p_{1/2}}=-9.14$           & $\varepsilon_{0f_{7/2}}=16.64$       \\
\multicolumn{1}{|c|}{}                          & $\varepsilon_{0g_{9/2}}=-7.24$           &                                  \\ \cline{1-3} 
\end{tabular}
\caption{\label{spentable} Experimentally deduced single-particle energy levels and model parameters utilized in the Hamiltonian for nuclei up to 6 nucleons above and below the 
$^{16}$O, $^{40}$Ca, and $^{56}$Ni cores.}
\end{table}

\subsection{Comparison to \textit{ab initio} results for $^{12}$C}
A recent paper \cite{DytrychMaris} reported \textit{ab initio} symmetry-adapted no-core shell model (SA-NCSM) calculations \cite{LauneyDD16} for the low-lying spectrum of $^{12}$C using the realistic nucleon-nucleon interaction JISP16 \cite{ShirokovMZVW07} for $\hbar\Omega=20$ MeV and $N_{\rm max}=8$ (or, including 10 harmonic oscillator major shells). The third $0^+$ state in the SA-NCSM calculations has been identified as the lowest $0^+$, $T=1$ state with excitation energy 21.42 MeV. This is consistent with the 18.16 MeV value calculated using our model (Fig. \ref{energyspectra}). Furthermore, the wave functions for the lowest isobaric analog  0$^{+}$  states in $^{12}$B, $^{12}$C, and $^{12}$N are expected to have very similar spatial parts, or deformation. Indeed, the SA-NCSM calculations reveal that  this 0$^{+}$ state in $^{12}$C is predominantly oblate with intrinsic spin 1, that is,  the $(\lambda\, \mu )=(1\,2)$ basis state contributes $\sim 61\%$ to this state, where $(\lambda\, \mu )$ are the deformation-related SU(3) quantum numbers \cite{CastanosDL88}. Exactly the same deformation dominates in the isobaric analog 0$^{+}$ state of $^{12}$N. The dominant features of these  isobaric analog 0$^{+}$ states in $A=12$  can be explained by strong pairing correlations (an isovector pair excitation given by the present model) as well as by strong collective modes, as suggested by the SA-NCSM. This is an interesting result pointing to the close interplay and overlap of pairing and deformation degrees of freedom, which has been also observed in other studies \cite{BahriED95, SviratchevaDV06,SviratchevaDV08}.

\subsection{Discrete derivatives and fine structure effects}
In this section a noteworthy test for the theory is implemented  and applied to the lowest isobaric analog $0^+$ states of $ee$ and $oo$ nuclei in the mass ranges 10 $\leq A \leq$ 22, 34 $\leq A \leq$ 46, and 50 $\leq A \leq$ 56. By considering the discrete derivatives of the energy function with respect to particle number, we are able to investigate the capability of the present model to reproduce fine features of nuclear dynamics. We utilize the formulae of  Ref. \cite{SviratchevaGeorgieva}, some of which are provided here for completeness, and follow the analysis reported in there. The discrete approximations of the $E_0$ energy are given as 
\begin{eqnarray}
Stg_{\delta }^{(m)}(x)
=&\frac{Stg_{\delta}^{(m-1)}(x+\frac{\delta  }{2})-Stg_{\delta
}^{(m-1)}(x-\frac{\delta }{2})}{\delta },\ m\geq 2, \nonumber \\
Stg_{\delta }^{(1)}(x)=&\left\{
\begin{array}{c}
\frac{E_0(x+\frac{\delta }{2})-E_0(x-\frac{\delta }{2})}{\delta },\ 
m\text{-even}
\\
\frac{E_0(x+\delta )-E_0(x)}{\delta },\ m\text{-odd},
\end{array}
\right.
\label{Stag_m}
\end{eqnarray}
where the variable $x=\{n,T_z,N_{+},N_{-}\}$ with $n$, $N_{+}$, and $N_{-}$ denoting the valence particles, valence protons and valence neutrons, respectively, and where increment $\delta \ge 1$. These approximations  (\ref{Stag_m}) eliminate  the large mean-field contributions (hence, often referred to as ``energy filters") and reveal the nuclear fine structure effects of pairing correlations. This is also true for the mixed derivatives, which are defined as
\begin{eqnarray} 
Stg_{\delta _1,\delta _2}^{(2)}(x,y) &=& \frac{E_0(x+ \delta _1,y+ \delta _2)-E_0(x+\delta _1,y)}{\delta _1 \delta _2} \\ \nonumber &-& \frac{E_0(x,y+\delta
_2)+E_0(x,y)}{\delta _1 \delta _2}, 
\label{Stag_2_mixed}
\end{eqnarray} 
where the variables
$(x,y)=\{n,T_z,N_{+1},N_{-1}\}$ and increments $\delta _{1,2} \ge 1$.    
We investigate different types of discrete derivatives of 
both the theoretical energies $E_0$ with their experimental counterparts, and analyze their staggering patterns. In our studies, $E_0$ is the energy plotted in Fig. \ref{energyspectra} with the Coulomb interaction removed. By removing the Coulomb interaction,  we isolate and study phenomena governed solely by the nuclear interaction.

\

As suggested in Refs. \cite{LauneyThesis,SviratchevaGeorgieva,KDL179,KDL180}, the significance of  various energy filters can be understood using phenomenological arguments that can be given a simple and useful graphical representation. Specifically, in the following subsections, each nucleus is represented by an inactive part, or a general $ee$ or $oo$ nucleus, schematically illustrated by a box,  $\square$, in which the interaction between the constituent particles does not change for a given energy filter. Active particles are represented by solid or empty dots for protons or neutrons, respectively, above the box. 

\subsubsection{Discrete derivatives with respect to the number of pairs and isospin projection: staggering  behavior and pairing gaps}

The description of $pn$ pairing correlations is crucial for reproducing staggering behavior and pairing gaps. The $Stg_{1}^{m}(T_z)$ and $Stg_{2}^{m}(2k)$ energy differences, $m = 1, 2, ...,$ isolate effects related to the various types of pairing in addition to changes in energy due to the different isospin values (symmetry term). We investigate these effects and provide insight into pairing correlations for $ee$ and $oo$ nuclei through analysis of the $Stg_{1}^{2}(T_z=0)$, $k$-odd and $Stg_{2}^{2}(2k)$, $T=1$ discrete derivatives in terms of the pairing gap relation 
\begin{eqnarray}
\tilde{\Delta }\equiv \Delta _{pp}+\Delta _{nn}-2\Delta _{pn} 
&\approx &\frac{1}{2}(\stackrel{\bullet \bullet }{\square }+\stackrel{\circ \circ }{\square }-2
\stackrel{\bullet \circ  }{\square }).
\label{deltatilde}
\end{eqnarray}
The result (\ref{deltatilde}) is a measure of the difference in the isovector pairing energy between $ee$ and $oo$ nuclei and follows from the well-known definition of the empirical like-particle pairing gap \cite{KDL85}

\begin{eqnarray} 
\Delta _{pp(nn)} &\equiv & \frac{1}{2}(BE(N_{+1}\pm 1,N_{-1}\mp 1) \\ \nonumber
 &-& BE(N_{+1}-1,N_{-1}-1)  \\ \nonumber
 &-& 2[BE(N_{\pm 1},N_{\mp 1}-1) \\ \nonumber &-& BE(N_{+1}-1,N_{-1}-1)])  \\ \nonumber
&=& \frac{1}{2}(\stackrel{\bullet \bullet }{\square }-\square -2[
\stackrel{\bullet }{\square }-\square]) \nonumber,
\label{ppnndelta}
\end{eqnarray}

which isolates the isovector pairing interaction of the $(N_{\pm 1})^{th}$ and $(N_{\pm 1}+1)^{th}$ protons (neutrons) for an even-even ($N_{+1}-1,N_{-1}-1$)-nucleus  (denoted by a square) \cite{KDL180}. As defined in \cite{LauneyThesis}, the $pn$ isovector pairing gap,
\begin{eqnarray}
\Delta _{pn} &\equiv & \frac{1}{2}(BE(N_{+1},N_{-1}) -BE(N_{+1},N_{-1}-1)  \\ \nonumber &-& [BE(N_{+1}-1,N_{-1}) \\ \nonumber &-& BE(N_{+1}-1,N_{-1}-1)]) \\ \nonumber 
&=& \frac{1}{2}(\stackrel{\bullet \circ }{\square }- \stackrel{\bullet
}{\square }
-[ \stackrel{\circ }{\square }-\square]), 
\label{gapPN}
\end{eqnarray}
is the pairing interaction of the $(N_{+1})^{th}$ proton and the
$(N_{-1})^{th}$ neutron. To correctly account for the $T=1$ mode of $pn$ pairing one should consider in Eq. (\ref{gapPN}) the $E_0$ energy of the $oo$ ($N_{+1},N_{-1}$) nucleus (that is, the energy of the isobaric analog state rather than its ground state energy, $BE$). For the remaining  $ee$ nuclei in Eq. (\ref{Stg2i}) replacing the symbol $E_0$ with $BE$ is justified.

\

For [$(k+T_z)$-even] and [$(k+T_z)$-odd] nuclei centered at $N=Z$ ($T_z=0$) and $N\ne Z$ ($T_z\ne 0$), the second-order discrete derivative
\begin{eqnarray}
Stg^{(2)}_{1}(T_z)&=&E_0(T_z+1)-2E_0(T_z) +  E_0(T_z-1), \nonumber \\  2k&=& \text{const},
\label{Stg2i}
\end{eqnarray}
can be written in terms of the pairing gap $\tilde{\Delta }$,
\begin{eqnarray} Stg^{(2)}_1 (T_z) \approx
\begin{cases}
2 \tilde{\Delta},\text{ } T_z=0,\text{ } k=\text{odd} \\
(-)^{(k+T_z)}\frac{4}{3(1+\delta_{T_z,0})}+V_r, \text{ otherwise},
\end{cases}
\label{Stg2i_int}
\end{eqnarray}
where in some cases the contribution from an additional residual nonpairing interaction $V_r$ cannot be fully removed. For $ee$ $N=Z$ nuclei, the additional $V_r$ term is a two-body interaction related to the nonpairing interaction of the three protons and three neutrons in $oo$ nuclei. However, for the $T_z\neq 0$ case of $ee$ and $oo$ nuclei the primary contribution of the residual interaction is from the symmetry energy. We also note that since $pp$, $nn$, and $pn$ $T=1$ pairs coexist \cite{KDL15,KDL45,SviratchevaGeorgieva}, $Stg^{(2)} _1 (T_z=0)$ does not simply account for the energy gained when two $pn$ pairs are created (in the first two $oo$ nuclei) and energy lost to destroy a $pp$ pair and a $nn$ pair in an $ee$ $N=Z$ nucleus. The relations (\ref{ppnndelta}-\ref{STG222N}) are based on the assumptions that the interaction of a particle within the box is independent of the type of added/removed particles and is the same for all protons (neutrons) above the box \cite{LauneyThesis}. 

\begin{figure*}[th]
\centerline{\epsfxsize=\textwidth \epsfbox{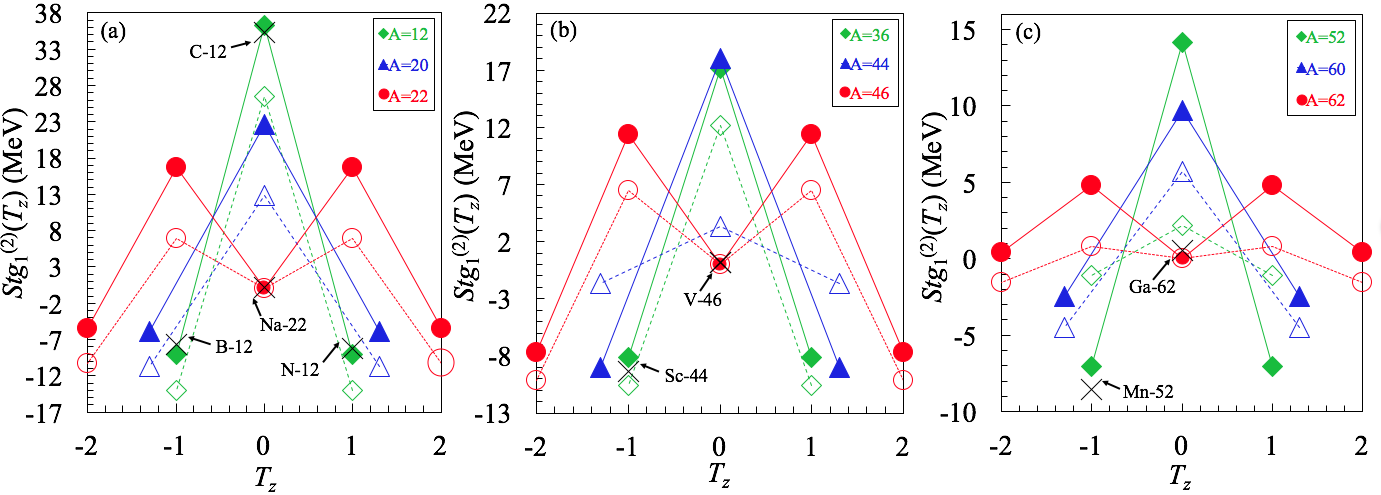}}
\caption{(Color online) Theoretical staggering amplitudes for the total energy (filled colored shapes) and the $pn$ and like-particle pairing energies (empty colored shapes) compared to experiment (black crosses) for (a)$^{16}$O; (b) $^{40}$Ca; (c) $^{56}$Ni core, as a function of the isospin projection $T_z$.}
\label{STG12abc}
\end{figure*}

We utilize $Stg_{1}^{2}(T_z=0)$ to isolate the effects related to like-particle and $pn$ pairing, which is described primarily by the symmetry term of our Hamiltonian. For example, in Fig. \ref{STG12abc}(a) the total energy and pairing energy contributions for $A=12$ are compared to experiment. Here, the symmetry energy contributes approximately 9 MeV to the total energy for $T_z=0$ and approximately 6 MeV for $T_z=\pm 1$, which highlights how crucial the isoscalar $T=0$ interaction is for reproducing the experimental energy in both Figs. \ref{STG12abc}(b) and (c). It is important to note the considerable differences in the energy ranges from Figs. \ref{STG12abc}(a-c). The large, yet gradually decreasing, energy differences from $^{16}$O to $^{56}$Ni may be attributed to the single-particle energy levels considered in $^{56}$Ni calculations, which are much closer in energy compared to those utilized for $^{16}$O and $^{40}$Ca.

\

\begin{figure}[th]
\centerline{\epsfxsize=3.5in\epsfbox{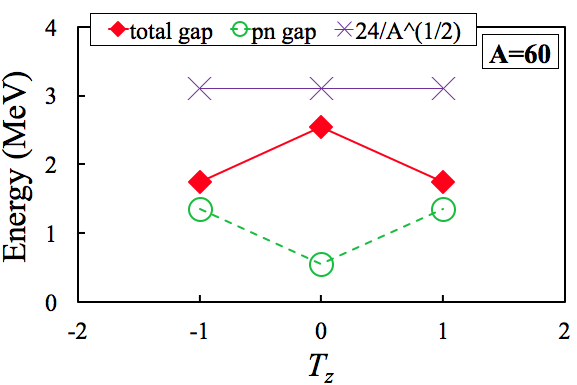}}
\caption{(Color online) Estimate for the total isovector pairing gap $\tilde{\Delta}$, $2 \Delta _{pn}$, and the empirical like-particle pairing gap $\Delta_{pp} + \Delta_{nn} = 24/\sqrt{A}$ for  $A=60$.}
\label{STG22abc}
\end{figure}

The second-order discrete derivative with respect to $2k$ (for a constant $T_z$),
\begin{eqnarray} 
Stg^{(2)} _2(2k)&=& \frac{E_0(2k+2)-2E_0(2k)+E_0(2k-2)}{4}, \nonumber \\ 
 &=& \frac{1}{4}(
\stackrel{\stackrel{\bullet}{\bullet} \stackrel{\circ}{\circ}}{\square
}-2\stackrel{\stackrel{\bullet}{}
\stackrel{\bullet }{ }}{\square }-\square),
\nonumber 
\label{Stg2n}
\end{eqnarray} 

 is related to the isovector pairing gap $\tilde{\Delta }$  \cite{SviratchevaGeorgieva},
 
\begin{eqnarray} 
Stg^{(2)} _2 (2k)\approx
    (-)^{(k+T_z)}\frac{\tilde{\Delta }}{3}+V_r, 
\label{STG222N}
\end{eqnarray} 

where in the $oo$ case $V_r$ is the nonpairing interaction of the last two protons with the last two neutrons in the ($2k+2$) nucleus.  The additional nonzero contribution of the symmetry energy prevents the isolation of the pairing gap relation $\tilde{\Delta}$ through Eq. (\ref{STG222N}). However, by using only the first two terms of the Hamiltonian (\ref{model}) in the calculations for $E_0$, we can eliminate the contribution of the symmetry energy. Hence, the staggering amplitude of the theoretical total pairing energy, which includes like-particle and $pn$ pairing energies, can provide an estimate of the $\tilde{\Delta }$ pairing gap using Eq. (\ref{STG222N}).  As an example, Fig. (\ref{STG22abc}) shows the total pairing gap for $A=60$ isotopes, which is estimated to be between 1.5-2.4 MeV. Since the approximation (\ref{ppnnapprox}) does not considerably fluctuate compared to the $pn$ pairing gaps with respect to $T_z$ \cite{SviratchevaGeorgieva}, we utilize the experimentally deduced like-particle pairing approximation,
\begin{equation}
\Delta_{pp}+\Delta_{nn}\approx \frac{24}{\sqrt{A}}.
\label{ppnnapprox}
\end{equation}

Using the total pairing gap (\ref{STG222N}) and its relation to the $pn$ and like-particle gap (\ref{deltatilde}), we provide an estimate for the $pn$ pairing gap $2 \Delta _{pn}$ that is between 0.5-1.5 MeV for $A=60$.  The like-particle pairing gap estimate, compared to the $pn$ gap, primarily contributes to the total gap for $A=60$. We note that in this staggering filter the single-particle term discontinuity may have an effect, and for lighter isotopes, where the energy difference between single-particle energies is larger, the effect is also larger.

\subsubsection{Discrete derivatives with respect to proton and neutron numbers: $N=Z$ irregularities}

\noindent As discussed in Ref. \cite{SviratchevaGeorgieva}, the second-order discrete mixed derivative $\delta V_{pn}(Z,N)$,
\begin{eqnarray}
\delta V_{pn}(Z,N)&=&\frac{E_0(Z+2,N+2)
-E_0(Z+2,N)}{4} \nonumber \\ &-&\frac{E_0(Z,N+2)+E_0(Z,N)}{4},
\label{dvpn}
\end{eqnarray}
\noindent represents, for even-even nuclei, the residual interaction between the last proton and the last neutron \cite{KDL179,KDL190}. It is well known that the attractive dip in the
$N=Z$ nuclei cannot be described by a model with an isovector interaction only. Hence, this filter is an important probe of
the $\alpha$-term in the model Hamiltonian (\ref{model}) that is related to $pn$ isoscalar interactions. 

\noindent Following the convention from the previous subsections, Eq. (\ref{dvpn}) can be graphically represented as
\begin{eqnarray}
\hspace{-0.2in}
\delta V_{pn}(Z,N)=\frac{1}{4}(
\stackrel{\stackrel{\bullet}{\bullet} \stackrel{\circ}{\circ}}{\square
}-\stackrel{\stackrel{\bullet}{}
\stackrel{\bullet }{ }}{\square }+
\stackrel{\stackrel{\circ}{} \stackrel{\circ }{}}{\square }-\square). \nonumber
\label{Stg2iGap}
\end{eqnarray}

\begin{figure}[th]
\centerline{\epsfxsize=3.5in\epsfbox{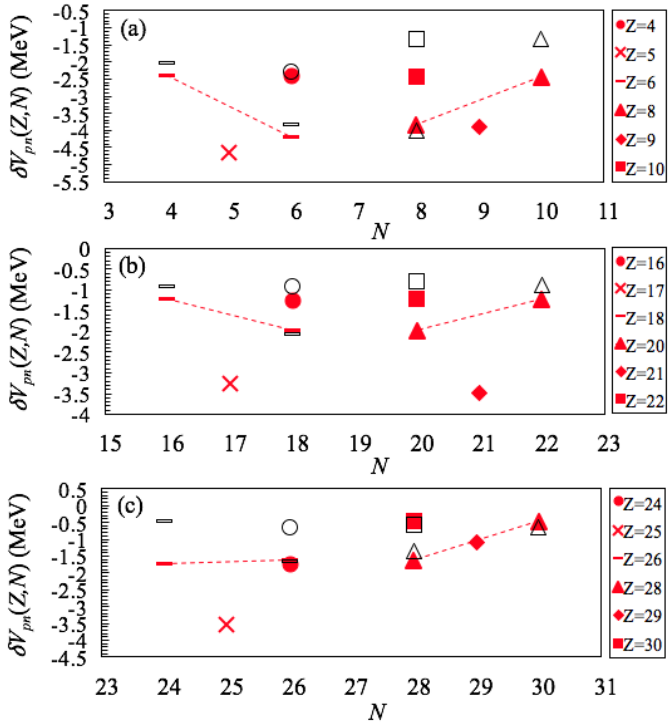}}
\caption{(Color online) The second-order discrete mixed derivative with respect to $Z$ and $N$,   shown  for (a) $4\le Z \le 10$, (b) $4\le Z \le 10$, and (c) $4\le Z \le 10$, where the filled colored shapes correspond to the theoretical calculations and the empty shapes correspond to their experimental counterparts. The energy filter eliminates the mean-field contribution from the energy and isolates the residual interaction between the last proton and last neutron in even-even nuclei; the $N=Z$ value is a probe for the $pn$ isoscalar interaction, shown here for (a) $^{12}$C and $^{16}$O, (b) $^{36}$Ar and $^{40}$Ca, (c) $^{52}$Fe and $^{56}$Ni.}
\label{DVpnabc}
\end{figure}

\noindent In contrast to the previous filters, the relation (\ref{dvpn}) does not display a consistent staggering pattern (Fig. \ref{DVpnabc}), but we expect that for fixed $Z$ there is a significant change in energy when $N=Z$. In this study, this filter can be applied to only selected nuclei, since the calculations are carried for up to 3 pairs. The model reproduces  experimentally deduced $N=Z$ values for the C, O, Ar, Ca, Fe, and Ni isotopes. With the exception of Fe ($Z=26$) the results agree remarkably well with the experimental data. The deviation may be as  a result of the absence of the $0d_{\frac{5}{2}+}$ single-particle energy level in our calculations, as described above. The good agreement points to the significance of the symmetry term in the model Hamiltonian (\ref{model}) and the physically relevant choice for the value of its strength $\alpha$ (Table \ref{spentable}).

\section{Conclusions}
We have presented a new shell model Hamiltonian that yields exact solutions for the lowest isobaric analog $J^\pi=0^+$, $T=0,..,3$ states that includes both like-particle and $pn$ pairing, as well as a symmetry term that is related to $pn$ isoscalar interactions. Adapted from Ref. \cite{SviratchevaGD_PRC04}, the model Hamiltonian utilizes experimentally deduced non-degenerate single-particles energies and includes an isoscalar $T=0$ interaction, which describes the interaction of nonpaired nucleons. The present results are based on the exact solutions for isovector pairing in non-degenerate single-particle energies  derived in Ref. \cite{PanDraayer}. The model utilizes only two adjustable parameters: the pairing strength, $G>0$, and $\alpha$, which is the strength of the isoscalar  interaction. We applied our model to even-$A$ nuclei for up to six particles above and below the $^{16}$O, $^{40}$Ca, and $^{56}$Ni cores and reported exact solutions for shell model $pp$, $nn$, and $pn$ pairing correlations for $ee$ and $oo$ nuclei in the mass ranges $10\leq A \leq 22$, $34\leq A \leq 46$, and $50\leq A \leq 62$. When comparing our results to a recent \textit{ab initio} study \cite{DytrychMaris} we found the same deformation dominates the isobaric analog 0$^{+}$ states in $^{12}$N, where the dominant features of these isobaric analog states in $A=12$ can be explained by both strong pairing correlations and strong collective modes. In addition to remarkably reproducing the energy spectra, we investigated how well the model captures fine features of nuclear dynamics by analyzing our results through discrete derivatives of the calculated energies. We isolated the effects related to like-particle and $pn$ pairing through theoretical staggering amplitudes for the total, $pn$, and like-particles energies. Estimates for the total isovector pairing gap and $pn$ contribution were provided for $A=60$, where the total gap is between 1.5-2.5 MeV and the $pn$ contribution is between 0.5-1.5 MeV. Additionally, the model correctly reproduces the $N=Z$ irregularity, which is a signature of non negligible isoscalar  $pn$ interaction, and we showed that the attractive dip expected for $N=Z$ nuclei was, indeed, well reproduced by the present results.

\section*{Acknowledgments}

We thank the National Science Foundation for supporting this work through the REU Site in Physics \& Astronomy (NSF grant 1262890) at Louisiana State University. 
This work was supported by the U.S. National Science Foundation (OIA-1738287, ACI -1713690), SURA, CUSTIPEN, and the National Natural Science Foundation of China (11675071). This work benefitted from computing resources provided by Blue Waters, LSU ({\tt www.hpc.lsu.edu}), and the National Energy Research Scientific
Computing Center (NERSC). The Blue Waters sustained-petascale computing project is supported by the National Science Foundation (awards OCI-0725070 and ACI-1238993) and the state of Illinois, and is a joint effort of the University of Illinois at Urbana-Champaign and its National Center for Supercomputing Applications. 
We also thank Grigor H. Sargsyan for providing SA-NCSM calculations for comparison. 

\section*{Appendix}
\appendix

\noindent The following equations are derived in Ref. \cite{PanDraayer} and are presented here for completeness, since different variables have been employed in the present numerical calculations.

\

$\textit{The $k=1$ case.} $ As defined in Ref. \cite{PanDraayer}, there is only one irreducible representation (irrep) [1,0,0] of the permutation group $S_{1}$. The eigenvalue for the $k=1$, $T=1$ case is given as
\begin{equation}
\label{eigenk1T1}
E_{\zeta}^{[1]T=1}= y^{(\zeta)} +2k\epsilon_{\rm avg },
\end{equation}
where the inverse spectral parameter $y^{(\zeta)}$ must satisfy 
\begin{equation}
\label{k1T1mod}
1+G\sum\limits_{j}\frac{\Omega_j}{y^{(\zeta)}-2\varepsilon_{j}}=0.
\end{equation}

$\textit{The k=2 case.}$ This case is solved for $T=0,2$ of the irrep [2,0,0] and $T=1$ of the irrep [1,1,0] of the permutation group $S_2$, where the eigenvalues for the $k=2$, $T=0,1,2$ cases are given as
\begin{equation}
\label{eigenenk2}
E_{\zeta}^{\substack{[2]T=2,0 \\ [1,1,0]T=1}}=y_{1}^{(\zeta)}+y_{2}^{(\zeta)}+2k\epsilon_{\text{avg}}.
\end{equation}
The inverse spectral parameters $y_{1}^{(\zeta)}$ and $y_{2}^{(\zeta)}$ for $T=2,0$ of [2,0,0] must simultaneously satisfy
\begin{equation}
\label{T2k2mod}
1+G\sum\limits_{j}\frac{\Omega_{j}}{y^{(\zeta)}_{i}-2\varepsilon_{j}}\pm \frac{2G}{y^{(\zeta)}_{m}-y^{(\zeta)}_{i}}=0,
\end{equation}
 where $i=1,2$, $m=2,1$, and $y_{1}^{(\zeta)} \ne y_{2}^{(\zeta)}$.

The inverse spectral parameters $y_{1}^{(\zeta)}$ and $y_{2}^{(\zeta)}$ of the $T=1$ case of the irrep [1,1,0] must simulataneously satisfy 
\begin{equation}
\label{T1k2mod}
1+G\sum\limits_{j}\frac{\Omega_{j}}{y^{(\zeta)}_{i}-2\varepsilon_{j}}=0,
\end{equation} for $i=1,2$, where $y^{(\zeta)}_{1}\ne y^{(\zeta)}_{2}$.

 $\textit{The k=3 case.}$ The irreps [3], [2,1,0], and [1$^{3}$] of the permutation group $S_3$ are solved for $T=3,1$, $T=2$, and $T=0$, respectively, where the eigenvalue equation for all $k=3$ cases is 
 \begin{equation}
 \label{eigenenk3}
E_{\zeta}^{\substack{[3]T=3,1 \\ [2,1,0]T=2 \\ [1^{3}]T=0}}=2k\epsilon_{\text{avg}}+\sum\limits_{i=1}^{3}y_{i}^{(\zeta)}.
 \end{equation}
 The inverse spectral parameters $y_{i}^{(\zeta)}$ for $i=1,2,3$ for the $T=3$ and $T=0$ cases of Eq. (\ref{eigenenk3}) must satisfy
 \begin{equation}
 \label{T3k3mod}
 1+G\sum\limits_{j}\frac{\Omega_{j}}{y_{i}^{(\zeta)}-2\varepsilon_{j}}-2G\sum_{\substack{m \\ i \ne m}}^{3}\frac{1}{y_{i}^{(\zeta)}-y_{m}^{(\zeta)}}=0
\end{equation}  
and
 \begin{equation}
 \label{T0k3mod}
 1+G\sum\limits_{j}\frac{\Omega_{j}}{y_{i}^{(\zeta)}-2\epsilon_{j}}=0,
\end{equation}  
respectively. The three resulting equations for both $T=3$ and $T=0$ when $i=1,2,3$ must be solved simultaneously, where  the solutions are only valid when $y^{(\zeta)}_{1} \neq y^{(\zeta)}_{2} \neq y^{(\zeta)}_{3}$, which is due to the antisymmetric nature of the wavefunction \cite{PanDraayer}.

\

For $T=2,1$ the inverse spectral parameters  $y_{i}^{(\zeta)}$ for $i=1,2,3$ must satisfy 
\begin{equation}
\label{T2k3mod}
1+G\sum\limits_{j}\frac{\Omega_{j}}{y_{i}^{(\zeta)}-2\varepsilon_j}-GF_{i}^{\substack{[2,1,0] \\ [3,0,0]}}(y_{1},y_{2},y_{3})=0,
\end{equation}
where, for simplicity, we introduce the relations $a=y_{1}^{(\zeta)}$, $b=y_{2}^{(\zeta)}$, and $c=y_{3}^{(\zeta)}$. For $T=2$ there are two sets of equations for $F_{i}^{[2,1,0]}(a,b,c)$ provided in \cite{PanDraayer} that yield solutions for $y_{i}^{(\zeta)}$ where $i=1,2,3$. The first set of equations,
\begin{subequations}
\label{T2k3modF1}
\begin{align}
R=\sqrt{a^{2}+b^{2}-bc+c^{2}-a(b+c)}\nonumber \\
F_{1}^{[21]}=-\frac{b+c-2a-R}{(a-b)(a-c)}, \\
F_{2}^{[21]}=\frac{a-2b+c-R}{(a-b)(b-c)}, \\
F_{3}^{[21]}=\frac{a+b-2c-R}{(a-c)(c-b)},
\end{align}
\end{subequations}
and the second set,
\begin{subequations}
\label{T2k3modF2}
\begin{align}
F_{1}^{[2,1,0]}=\frac{b+c-2a-R}{(a-b)(a-c)}, \\
F_{2}^{[2,1,0]}=\frac{a-2b+c+R}{(a-b)(b-c)}, \\
F_{3}^{[2,1,0]}=\frac{a+b-2c+R}{(a-c)(c-b)},
\end{align}
\end{subequations}

both produce solutions for (\ref{T2k3mod}). The solutions provided by Eq. (\ref{T2k3modF1}) are only valid when $y^{(\zeta)}_{1}=y^{(\zeta)}_{2}<y^{(\zeta)}_{3}$, $y^{(\zeta)}_{1}=y^{(\zeta)}_{3}<y^{(\zeta)}_{2}$, and $y^{(\zeta)}_{2}=y^{(\zeta)}_{3}<y^{(\zeta)}_{1}$, and solutions provided by Eq. (\ref{T2k3modF2}) are only valid when $y^{(\zeta)}_{2}=y^{(\zeta)}_{1}>y^{(\zeta)}_{3}$, $y^{(\zeta)}_{3}=y^{(\zeta)}_{1}>y^{(\zeta)}_{2}$, and $y^{(\zeta)}_{3}=y^{(\zeta)}_{2}>y^{(\zeta)}_{1}$.

\

The most complicated case for $k=3$ is $T=1.$ The equations for this case derived in \cite{PanDraayer} include a $\gamma$-term and an $\alpha$-term that are not present in our adjusted equations, for we substituted $\gamma=1$ and the $\alpha$ term. The set of equations for $F_{i}^{[3,0,0]}(a,b,c)$ is 
\begin{subequations}
\label{T1k3modF}
\begin{align}
F_{1}^{[3,0,0]}=\frac{b\beta-2c(1+\beta)+a(2+\beta)}{(a-b)(a-c)(1+\beta)}, \\ 
F_{2}^{[3,0,0]}=-\frac{a+b+2b\beta-2c(1+\beta)}{(a-b)(b-c)(1+\beta)}, \\
F_{3}^{[3,0,0]}=\frac{-a+c-b\beta+c\beta}{(a-c)(c-b)(1+\beta)},
\end{align}
\end{subequations}
where the relations for $\beta$ that produce solutions are

\begin{eqnarray}
\label{T1k3B1}
\beta_1&=&\frac{1}{9 (a - c) (-b + c)}\bigg(2 a^2 - 3 b^2 + 4 a (b - 2 c) \\  \nonumber &+& 2 b c + 3 c^2- \bigg(\frac{h_3}{(h_1 + h_2)^{1/3}}+ (h_1 + h_2)^{1/3}\bigg)\bigg), 
\\
\label{T1k3B2}
\beta_2&=&-\frac{1}{36(a-c)(c-b)}\bigg(h_4-\frac{2(\sqrt{3}i+1)h_5}{(h_1+h_2)^{1/3}} \\ \nonumber &+&2(-1+\sqrt{3}i)(h_1+h_2)^{1/3}\bigg), 
\\
\label{T1k3B3}
\beta_3&=&-\frac{1}{36(a-c)(c-b)}\bigg(h_4-\frac{2(\sqrt{3}i-1)h_5}{(h_1+h_2)^{1/3}} \\ \nonumber
&-&2(1+\sqrt{3}i)(h_1+h_2)^{1/3}\bigg).
\end{eqnarray}
The arguments $h_{1,..,5}$ in (\ref{T1k3B1}-\ref{T1k3B3}) are 

\begin{center}
\begin{eqnarray}
\label{h1}
\nonumber
h_{1}=-9(a-b)(a-c)(b-c)\sqrt{3D}
\end{eqnarray}

\begin{eqnarray}
\label{D}
D&=&-9 a^6 + 27 a^5 b - 79 a^4 b^2 + 113 a^3 b^3 - 79 a^2 b^4 \nonumber \\ \nonumber &+& 27 a b^5 - 9 b^6 + 27 a^5 c + 23 a^4 b c - 23 a^3 b^2 c - 23 a^2 b^3 c \\ \nonumber &+& 23 a b^4 c + 27 b^5 c- 79 a^4 c^2 - 23 a^3 b c^2 + 69 a^2 b^2 c^2 \\ \nonumber &-& 23 a b^3 c^2 - 79 b^4 c^2+ 113 a^3 c^3 - 23 a^2 b c^3 - 23 a b^2 c^3 \\ \nonumber &+& 113 b^3 c^3  - 79 a^2 c^4 + 23 a b c^4 - 79 b^2 c^4 + 27 a c^5 \\ \nonumber &+& 27 b c^5 - 9 c^6
\end{eqnarray}
 
\begin{eqnarray}
\label{h2}
 h_{2}&=&-8a^6+33a^5b-6a^4b^2+53a^3b^3 + 144 a^2 b^4 \nonumber \\ \nonumber &-&  
 27ab^5+27b^6+15a^5c - 153 a^4bc- 135 a^3b^2c \\ \nonumber &-& 735 a^2 b^3 c - 153 a b^4c - 135b^5 c + 
 39a^4 c^2 + 441a^3bc^2 \\ \nonumber &+& 1305a^2 b^2 c^2 + 1041ab^3c^2 + 414b^4 c^2 - 199a^3c^3 \\ \nonumber &-& 1311a^2bc^3- 1911ab^2c^3- 899b^3c^3+477a^2c^4 \\ \nonumber &+& 1611abc^4+ 1152b^2c^4 - 513ac^5-783bc^5+216c^6
\end{eqnarray}

\begin{eqnarray}
\label{h3}
\nonumber
 h_{3}&=& 4a^4 - 11 a^3 b + 40 a^2 b^2 - 6 a b^3 + 9 b^4  \\ \nonumber &-& 5 a^3 c - 47 a^2 bc - 62 a b^2 c- 30 b^3c+ 31 a^2 c^2 \\ \nonumber &+& 109 a b c^2 + 76 b^2 c^2 - 57 a c^3 - 87 b c^3 + 36 c^4
\end{eqnarray}
 
\begin{eqnarray}
\label{h4}
\nonumber
 h_{4}=-4(2 a^2 - 3 b^2 + 4 a (b - 2 c) + 2 b c + 3 c^2)
\end{eqnarray}

\begin{eqnarray}
\label{h5}
\nonumber
h_{5}&=& 4 a^4 - 11 a^3 b + 40 a^2 b^2 - 6 a b^3 + 9 b^4 \\ \nonumber &-& 5 a^3 c - 47 a^2 b c - 62 a b^2 c - 30 b^3 c + 31 a^2 c^2 \\ \nonumber &+& 109 a b c^2 + 76 b^2 c^2 - 57 a c^3 - 87 b c^3 + 36 c^4
\end{eqnarray}
\end{center}

\bibliographystyle{apsrev}
\bibliography{staggeringSp4bib}
\end{document}